 \documentclass[twocolumn,preprint]{aastex63}

\usepackage{microtype,longtable,verbatim,float,booktabs,graphicx}
\usepackage{tabularx}
\usepackage{amsmath} 
\usepackage{textcomp}
\newcommand{\angstrom}{\textup{\AA}}

\newcommand{\SFinf}{$\rm SF_{\infty}$}
\newcommand{\redchi}{$\chi_{\nu}^2  \;$}
\newcommand{\Eddratio}{$L_{\rm bol}/L_{\rm Edd}$}

\received{\today}
\revised{---}
\accepted{---}

\submitjournal{ApJ}

\shorttitle{Binary SMBHs in LSST DDFs}
\shortauthors{Davis et al.}

\begin{document}

\title{Reliable Identification of Binary Supermassive Black Holes \\ from Rubin Observatory Time-Domain Monitoring}

\correspondingauthor{Megan C. Davis}
\email{megan.c.davis@uconn.edu}

\author[0000-0001-9776-9227]{Megan C. Davis}
\altaffiliation{NSF Graduate Research Fellow}
\affiliation{Department of Physics, 196A Auditorium Road, Unit 3046, University of Connecticut, Storrs, CT 06269, USA}

\author[0009-0001-6853-9470]{Kaylee E. Grace}
\affiliation{Department of Physics \& Astronomy, 217 Sharp Lab, Newark, DE 19716, USA}

\author[0000-0002-1410-0470]{Jonathan R. Trump}
\affiliation{Department of Physics, 196A Auditorium Road, Unit 3046, University of Connecticut, Storrs, CT 06269, USA}

\author[0000-0001-8557-2822]{Jessie C. Runnoe}
\affiliation{Department of Physics and Astronomy, Vanderbilt University, Nashville, TN 37235, USA}
\affiliation{Department of Life and Physical Sciences, Fisk University, 1000 17th Avenue N, Nashville, TN, 37208, USA}

\author[0000-0001-8688-5273]{Amelia Henkel}
\affiliation{Department of Physics, University of New Hampshire, 9 Library Way, Durham,
NH 03824, USA}


\author[0000-0002-2183-1087]{Laura Blecha}
\affiliation{Department of Physics, University of Florida, Gainesville, FL 32611, USA}

\author[0000-0002-0167-2453]{W. N. Brandt}
\affiliation{Department of Astronomy \& Astrophysics, 525 Davey Lab, The Pennsylvania State University, University Park, PA 16802, USA}
\affiliation{Institute for Gravitation and the Cosmos, The Pennsylvania State University, University Park, PA 16802, USA}
\affiliation{Department of Physics, 104 Davey Lab, The Pennsylvania State University, University Park, PA 16802, USA}

\author[0000-0002-5557-4007]{J. Andrew Casey-Clyde}
\affiliation{Department of Physics, 196A Auditorium Road, Unit 3046, University of Connecticut, Storrs, CT 06269, USA}

\author[0000-0003-3579-2522]{Maria Charisi}
\affiliation{Department of Physics and Astronomy, Vanderbilt University, Nashville, TN 37235, USA}



\author[0000-0002-6020-9274]{Caitlin Witt}
\affiliation{Center for Interdisciplinary Exploration and Research in Astrophysics (CIERA), Northwestern University, Evanston, IL 60208, USA}
\affiliation{Adler Planetarium, 1300 S. DuSable Lake Shore Dr., Chicago, IL 60605, USA}


\begin{abstract}
Periodic signatures in time-domain observations of quasars have been used to search for binary supermassive black holes. These searches,
across existing time-domain surveys, have produced several hundred candidates. The general stochastic variability of quasars, however, can masquerade as a false-positive periodic signal, especially when monitoring cadence and duration are limited. In this work, we predict the detectability of binary supermassive black holes in the upcoming Rubin Observatory Legacy Survey of Space and Time (LSST). We apply computationally inexpensive sinusoidal curve fits to millions of simulated LSST Deep Drilling Field light curves of both single, isolated quasars and binary quasars. Period and phase of simulated
binary signals can generally be disentangled from quasar variability.
Binary amplitude is overestimated and poorly recovered for two-thirds of potential binaries due to quasar accretion variability. Quasars with strong intrinsic variability
can obscure a binary signal too much for recovery. We also find that the most luminous quasars
mimic current binary candidate light curves and their properties: false positive rates are 60\% for these quasars. 
The reliable recovery of binary period and phase for a wide range of input binary LSST light curves is promising for multi-messenger characterization of binary supermassive black holes. However, pure electromagnetic detections of binaries using photometric periodicity with amplitude greater than 0.1 magnitude will result in samples that are overwhelmed by false positives. This paper represents an important and computationally inexpensive way forward for understanding the true and false positive rates for binary candidates identified by Rubin.

\end{abstract}


\section{Introduction}\label{sec:intro}

Binary supermassive black holes (SMBHs) are an expected consequence of galaxy mergers (e.g., \citealp{KormendyHo2013}). Close, sub-parsec binary SMBHs should be the loudest gravitational wave sources in the Universe, likely producing the recently announced low-frequency gravitational wave background detection from the North American Nanohertz Observatory for Gravitational Waves (NANOGrav; \citealp{2023Nanograv15yrEvidence, 2023nanograv15yearBSMBHConstraints}) and other global Pulsar Timing Arrays (PTAs; \citealp{Antoniadis2023secondDR, Reardon2023}). 
The binary SMBH population traces galaxy merger rates, helping our understanding of both SMBH-galaxy co-evolution and the hierarchical structure formation of galaxies \citep{Comerford}. Constraints on inferred merger rates and the gravitational wave background indicate false-positive contamination of the binary quasar candidate population (see binary candidate discussion below; \citealp{Sesana_2018}).
Identifying binary SMBHs via electromagnetic methods makes individual gravitational wave sources easier to detect with pulsar timing arrays (e.g., \citealp{Arzoumanian20203C66B, LiuVigeland2021}) and the upcoming, revolutionary space-based gravitational wave observatory, LISA \citep{LISA, LISAWhitePaper2023}. Beyond gravitational wave astrophysics, an electromagnetic binary SMBH detection further opens the door to study black hole inspiral and accretion environments, cosmology, and more \citep{De_Rosa_2019}.  
 
There are confirmed electromagnetic detections of dual quasars that are closely separated (embedded within the same galaxy via merger, kiloparsec to parsec scales) but not yet gravitationally bound (parsec to sub-parsec scales) as binary SMBHs (e.g.,\citealp{VanWassenhove2012,BlumenthalBarnes2018,De_Rosa_2019,Chen2022Dual,DualAGN_2023}). 
Candidate binary SMBH systems are identified
from observations spanning the electromagnetic spectrum, such as notable radio interferometry candidates in \cite{Rodriguez2006} and \cite{Bansal2017}. 
In this work, we focus on photometric searches of binary SMBHs in the optical regime. 

Binary SMBH systems are expected to exhibit
periodic modulation of photometric light curves due to Doppler boosting or periodic accretion \citep[more detailed descriptions of these binary models can be found in Section \ref{sec:binarysig};][]{Charisi_2022}.
Equal-mass binaries (of mass ratio within a factor of a few) will generally have the strongest periodic signatures in photometric observations, in addition to being the loudest expected gravitational wave sources \citep{D_Orazio_2013}.
\cite{Graham2015, Charisi2016, Liu2019,Chen2020, Chen2022b} are all examples of optical photometric searches for periodic behavior that have produced several hundred binary SMBH candidates (see \citealp{De_Rosa_2019, DOrazioCharisiBook2023} and references therein for a review). Despite these efforts, we have yet to confirm these binary SMBH candidates and true binaries remain elusive.  A strongly varying single quasar can appear indistinguishable from a binary SMBH system because intrinsic optical quasar variability can masquerade as a periodic signature of a binary SMBH. Limited survey cadence and duration can especially lead to false positive detections \citep{Vaughan2016, Barth_2018, Witt_2022}.

Upcoming massive time-domain surveys, like the Vera Rubin Observatory (Rubin)'s 10-year Legacy Survey of Space and Time (LSST), will open a new frontier for electromagnetic detection of binary SMBHs. LSST is expected to observe between 20-100 million active galactic nuclei (AGN), with time-domain monitoring that has a cadence and duration appropriate to detect potential sub-parsec binary SMBHs \citep{Ivezic2017, Kelley2019c, Kelley2021, XinHaiman2021, Charisi_2022}. These AGN will be observed via the main LSST survey, Wide-Fast-Deep (WFD; a footprint of roughly 18,000~deg$^2$), that uses 80\%-90\% of the available observation time and the Deep Drilling Fields (DDFs; each field has a footprint of roughly 9.6~deg$^2$), which will use a portion of the remaining 10\%-20\% of the observation time \citep{Bianco_2022LSST}.
In this work we focus on photometric light curves in DDF observations, representing the highest-cadence and deepest photometry for time-domain observations of AGN by Rubin/LSST.
The DDFs are expected to observe over 40,000 AGN \citep{brandt2018LSSTwhitepaper}.


Rubin's LSST, our best chance at identifying binary SMBHs with electromagnetic observations, also pushes us further into the era of big data, as it is predicted that it will produce over 20 terabytes of data per night\footnote{https://www.lsst.org/about/dm (\today)}. 
In order to triage data as it arrives, LSST will have to employ pipeline-style brokerage, or target-type sorting, using various, computationally inexpensive analysis methods, from curve fitting to machine learning \citep{Saha2016ANTARES,Narayan2018AlertBrokers, Matheson2021ANTARES}. In this work, we test a computationally inexpensive method for identifying binary SMBHs using simulated periodic and single-quasar light curves that resemble the expected observational properties of Rubin's LSST. This method, simple sinusoidal curve fits, could work as an initial filter to identify binary candidates for further, more intense follow-up like that of \cite{Witt_2022}. \cite{Witt_2022} applies complex Bayesian methods to simulated CRTS and LSST light curves that require several CPU hours per source, which is computationally prohibitive to apply to the entire database.
Our work can predict LSST detection rates for binary SMBHs and give insight into what types of quasars will produce false detections from millions of light curves within a few hours.

In the next section, Section \ref{sec:datgen}, we describe the construction and underlying models of the Rubin LSST light curves. Section \ref{sect:Analysis} details the simple sinusoidal curve fitting. We present the results of the light curve fitting, including predicted false positive rates and population statistics, in Section \ref{sec: results}. We discuss the results, with recommendations for future Rubin LSST observations and comparisons with gravitational wave searches, in Section \ref{sec: discuss}. We summarize our conclusions and outline future prospects in Section \ref{sec:Conclusion}.

We assume a flat, $\Lambda$CDM cosmology
with $\Omega_\Lambda$ = 0.7, $\Omega_M$ = 0.3, and $H_0$ = 67.4 $\pm$ 0.5 km s$^{-1}$ Mpc$^{-1}$ for this work \citep{Planck2018}.

\section{Light Curve Generation} \label{sec:datgen}
\begin{figure*}[t]
\centering
\noindent\includegraphics[width=\textwidth]{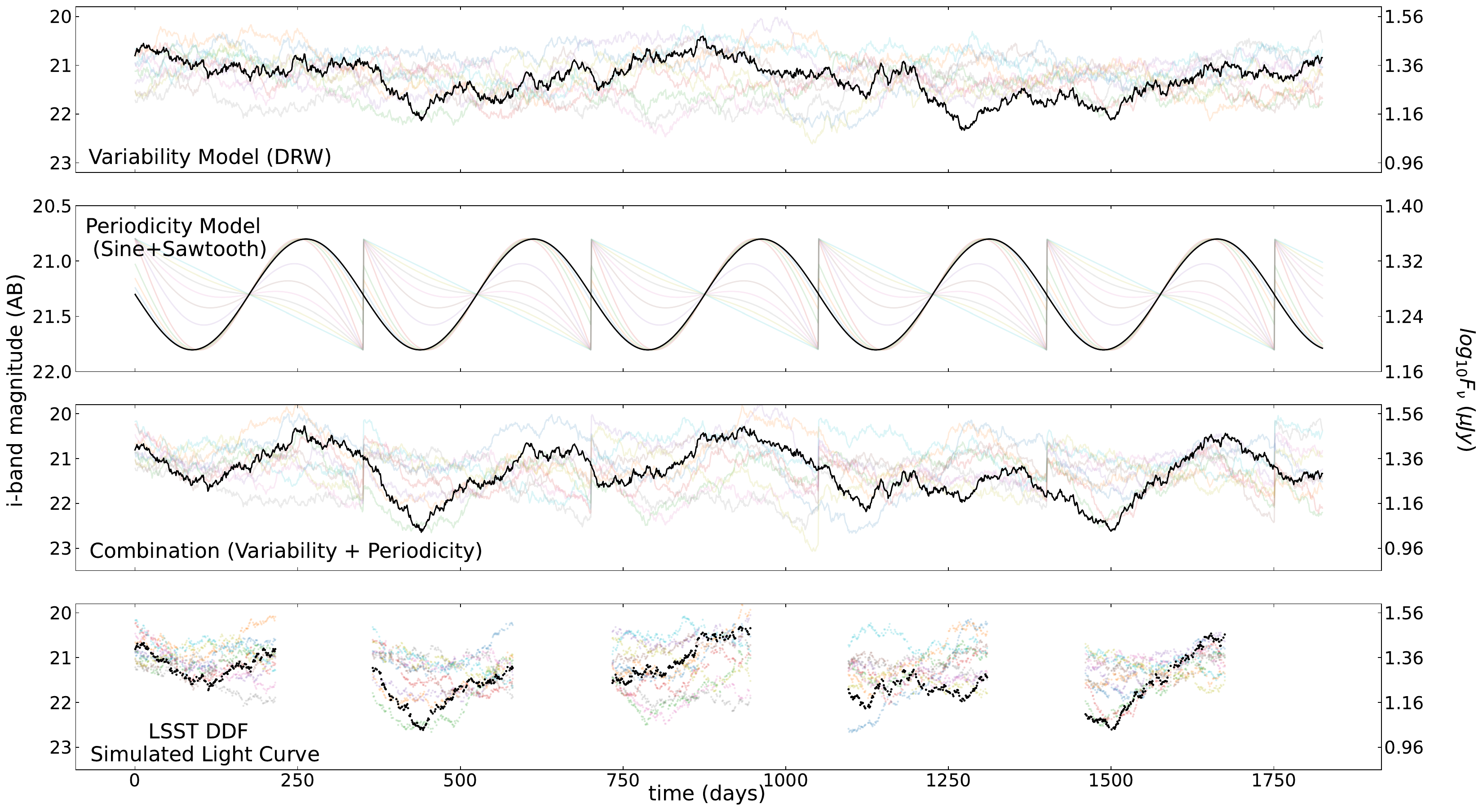}
\caption{A summary of the light curve assembly as outlined in Section \ref{sec:datgen}. The top most panel is the variability model/base light curve creation for a $10^8$ $M_{\odot}$ quasar at z $\sim$ 1 that is accreting at 10\% of its Eddington limit, with 10 additional realizations in faded colors behind the main black continuous line. The second panel reveals the periodicity model of the binary (some combination of sine and sawtooth waves). A black smooth sine is depicted here with 10 additional realizations behind it (in 10\% increments of pure sine to sawtooth). A light curve for an isolated, single quasar would skip the second and third panel steps. We add the two models together in the third panel. We finally then add noise and apply a cadence and seasonal duration to match the expected Rubin LSST DDF observation in the final panel. }\label{fig:LCassembly}
\end{figure*}

Our goal is to create realistic light curves of quasars, both isolated (single-SMBH) and binary systems, for Rubin's LSST Deep Drilling Fields (DDFs).
Light curve generation is accomplished in three steps: 1.\ generating the quasar variability, 2.\ creating the binary signal (for binaries only), and then 3.\ customizing for the observational noise, cadence, and monitoring duration specific to the LSST DDFs in the $i$-band. Figure \ref{fig:LCassembly} summarizes these steps, with the details in the sections that follow. Both isolated and binary quasars are expected to have accretion disks that exhibit stochastic variability, as close-separation binaries are predicted to have a circumbinary accretion disk and mini-disks around each SMBH \citep{Charisi_2022, BoganovicReview2022, DOrazioCharisiBook2023}. This is shown in the first panel of Figure \ref{fig:LCassembly}.
The binary systems are expected to have an additional periodic variability component, as described in Section \ref{sec:binarysig} and seen in panel two of Figure \ref{fig:LCassembly}. The customization of the light curves included implementing the expected cadence, duration, and noise properties of the LSST DDFs, as laid out in Section \ref{sec:RubinLSSTLCs} and seen in the last panel of Figure \ref{fig:LCassembly}.
The range of quasar and binary parameters used to create the light curves are shown in Table \ref{tab:paramgrid}.

\begin{table}[t]
\begin{center}
\begin{tabular}{ |c|c|c| } 
\hline
Parameter & [lower bound, upper bound] \\
\hline
log($M/M_{\odot}$) & [6.5, 9.5]  \\ 
log($L/L_{edd}$) & [-2, 0]  \\ 
z & [0.1, 2.5] \\ 
amplitude (mag) & [0.0, 1.0] \\ 
period (days) & [360,  1080]  \\ 
sawtoothiness (\%) & [0.0, 1.0 ]\\ 
phase &  [0.0, 2$\pi$]\\
\hline
\end{tabular}
\end{center}
\caption{The range of quasar properties of total binary mass, luminosity, and redshift and binary signal parameters of amplitude, period, signal type, phase used to generate the simulated light curves. A zero amplitude binary signal represents the non-binary (single) quasar population. We sample each parameter from a uniform distribution within its stated range.}
\label{tab:paramgrid}
\end{table}

In order to explore the full parameter range and unique combinations of their respective uniform and continuously sampled ranges, we generated a total of 4,812,500 $i$-band LSST light curves. Over 3.6 million light curves met a $i$-band magnitude limit of 23.5 that represents a light curve with sufficient signal-to-noise for reliable fitting, and they were kept for the analysis described in Section \ref{sect:Analysis}.

\subsection{Quasar Properties}\label{sec:QuasProps}

Our goal is to simulate light curves for a broad and representative range of the quasar population to be observed by Rubin.
We made a grid of quasar properties spanning black hole mass (total mass of the binary, in the case of a binary SMBH light curve), luminosity, and redshift, as seen as the first three entries of Table \ref{tab:paramgrid}. These ranges of quasar properties are representative of the quasar population observed by the Sloan Digital Sky Survey and other large surveys
(e.g. \citealp{Kollmeier_2006, Vestergaard_2009, Kelly_2010, Shen_2011, Wu_2022}). We chose this redshift range, specifically, because a majority of $i < 23.5$ of DDF quasars will be in a redshift range of $z < 2.5$ \citep[see Table 10.2 of Chapter 10 in][]{lsstsciencecollaboration2009lsst}. This restriction will also avoid 
Ly$\alpha$ forest absorption
in the bluest LSST filter and is further motivated by GW searches of binary candidates \citep{Charisi_2022}. The parameters are uniformly randomly sampled from within each bound to approximate continuous distributions. 
Having such a broad range and thousands of unique combinations of quasar parameters allows us to probe the properties of detectable binaries and of false-positive isolated SMBHs with this method. 

Our light curves do not include 
host-galaxy contamination, which may be important for low luminosity AGN within the DDF footprints \citep{lsstsciencecollaboration2009lsst}. Host-galaxy light is constant and will dilute the fractional variability observed, equally for binary and variability components. It would not change the relative variability of DRW compared to binary signals. 
For this reason,
we predict that the addition of host-galaxy light
will not change our general conclusions about binary identification and false-positive rates relative to DRW variability contribution.

\subsubsection{Variability Model}
The UV/optical variability of a quasar is effectively described as a stochastic process specifically in the form of a damped random walk (DRW) \citep{Kelly2009,MacLeod_2010,MacLeod2012,Zu_2013}. This variability is likely driven by thermal fluctuations in the accretion disk \citep{Kelly2009,MacLeod_2010, Kozlowski2010, Zu_2013, Suberlak2021}. The DRW structure function, SF, is defined as: 
    \begin{equation}
        \rm SF \left ( \Delta t \right) = SF_{\infty} \left( 1 - e^{-|\Delta t |/  \tau}\right )^{1/2}
    \end{equation}
The two parameters of the DRW are the characteristic damping timescale, $\tau$, and DRW amplitude, \SFinf. The values of each parameter depend on rest-frame wavelength, quasar luminosity, black hole mass, and redshift \citep[e.g.,][]{Uomoto1976,Hook1994, Giveon1999, Hawkins2002, VandenBerk2004,deVries2005,Wilhite2008, Bauer2009, Kelly2009, Kozlowski2010, MacLeod_2010, MacLeod2012}. We chose to describe them via Equation 7 of \cite{MacLeod_2010}:
    \begin{equation}
    \begin{aligned}
         \log f = A + B \log \left( \frac{\lambda_{\rm RF}}{4000 \angstrom} \right) + C \left( M_i + 23 \right) \\
         + D \log \left( \frac{M_{\rm BH}}{10^9 M_{\odot}} \right) + E \log \left ( 1 + z \right)
    \end{aligned}
    \end{equation}
where $\lambda_{\rm RF}$ is the rest-frame wavelength, $M_i$ is the absolute $i$-band magnitude of the target, $M_{\rm BH}$ is the mass of the black hole, and $z$ is redshift. We used the equation above with the best-fit coefficients of \citet{MacLeod_2010, MacLeod2012} summarized in Table \ref{tab:DRWparams}, for quasars with black hole mass, luminosity, and redshift drawn from the ranges described in Table \ref{tab:paramgrid}. We use the \texttt{generate\_dampedRW} function of the \texttt{AstroML} python package to create each DRW from the coefficients calculated by Equation 2. 

The DRW is an effective empirical model for quasar variability on timescales of months to years (although it appears to be less effective on shorter and longer timescales, \citealp{Mushotzky_2011,KellyCARMA2014,Vaughan2016,Kozlowski_2017a, Moreno2019,Kozlowski_2021, YuRichards2022}). We chose the DRW for its simplicity and accuracy in describing the variability of quasar accretion disks with the parameter space described above. We note, however, that the damping timescale is not well-constrained
by observations for much of the quasar parameter space \citep{MacLeod_2010, MacLeod2012, Kozlowski_2017a, Kozlowski_2017b}. The damping timescale $\tau$ is likely to drive the ``recovered" periods of false-positive signals. There is also some debate in the literature about the details of the correlations of the DRW parameters and quasar properties \citep[see Table 4 in][]{Suberlak2021}, although a detailed analysis by \cite{Suberlak2021} found similar correlations to \cite{MacLeod_2010, MacLeod2012}. We use the best-available model for $\tau$ in our simulated light curves, while noting that future long-duration monitoring (e.g., with Rubin/LSST) is likely to improve the quantitative description of quasar variability on long timescales.

The general schema of this project is flexible enough to incorporate other variability models in future work, like the  continuous-time autoregressive moving average (CARMA) models \citep{KellyCARMA2014} and the damped harmonic oscillator (DHO) model \citep{YuRichards2022}. Future work can also include a more detailed consideration of the differences between accretion disks around single quasars with the circumbinary and mini-disks in binary quasar \citep{ArtymowiczLubow1996, MacFadyen_2008, WS_2023}.

\begin{table}[t]
\begin{center}
\begin{tabular}{ |c|c|c|c|c|c| } 
\hline
f & A & B & C & D & E\\
\hline
\SFinf [mag] & -0.36 & -0.479 & 0.131 & 0.18 & 0.0 \\ 
$\tau$ [days] & 2.7 & 0.17 & 0.03 & 0.21 & 0.0 \\ 

\hline
\end{tabular}
\end{center}
\caption{The coefficients of the best-fit DRW parameter formulae from \cite{MacLeod_2010}, with updates on the A coefficient from \cite{MacLeod2012} to account for long-term variability.}
\label{tab:DRWparams}
\end{table}

\subsection{Binary Signal Parameters}\label{sec:binarysig}
 
We describe the electromagnetic signal of a binary quasar with a semi-sinusoidal periodic function that is parameterized with amplitude, period, phase, and ``sawtoothiness.'' The ranges of amplitude and period are shown in Table \ref{tab:paramgrid} and are designed to include close-separation binaries that have values detectable within the first half of the LSST DDF monitoring duration, but not so short than the binary evolves within the duration of the light curve. We sample phase from the full range of [0, 2$\pi$]: phase is generally treated as a nuisance parameter in electromagnetic binary searches, although its recovery is important for gravitational wave detection \citep{Charisi_2022}.
We additionally allowed for an extra parameter (``sawtoothiness"/``burstiness") to account for non-sinusoidal, sudden variability and allow for combinations of both functions.

Sinusoidal variability is predicted for binary SMBHs from relativistic Doppler boosting \citep{D_Orazio_2015,Charisi_2018,De_Rosa_2019,Charisi_2022}. Especially at the small separations associated with gravitational wave emission, relativistic Doppler boosting will cause periodic modulation of the photometric light curve in the form of a sine wave if the orbits are circular \citep{D_Orazio_2015b,Tang_2018}. This is the most ideal binary signature as all of the binary signal parameters (amplitude, period, and phase) can be directly linked to the binary's geometry and orbital dynamics, and, thus, to the gravitational wave signal counterpart \citep{Charisi_2022}. The photometric binary amplitude is related to the orbital velocity of the binary in the Doppler Boost model. The amplitude has uncertain connections to binary parameters for other binary models \citep{duffell2020}. We sample periodic binary amplitude from a broad, generic range of magnitudes to test general recovery with Rubin. Our model for the binary signal is empirical rather than explicitly connected to binary mass or luminosity ratio, although our underlying assumptions are generally associated with equal-mass binaries with an active secondary SMBH, since these binary systems are expected to produce the strongest periodic signatures in photometry \citep{D_Orazio_2013, DOrazioCharisiBook2023}.

The ``bursty'' sawtooth can be tied to eccentric orbits and to orbital periods of hot spots (overdensities) on the edge of the internal cavity of the circumbinary accretion disk and the sudden inflow of that built-up material \citep[periodic accretion;][]{duffell2020}. This is the less ideal binary signal as it is more difficult to tie the amplitude, period, and phase back to the orbital dynamics of the binary
\citep{Charisi_2022}. Hot spots in the inner edge of the circumbinary accretion disk can cause a photometric period that is 5-8 times the actual orbital period for equal-mass binaries \citep{DOrazio_2013, Farris_2014, Miranda2017, Charisi_2022}. We do not explicitly model longer observed-frame period sawtooth waves than sine waves nor do we impose a higher frequency sinusoid on top of sawtooth \citep{duffell2020}. Instead our range of binary parameters is defined to include a representative range of binaries detectable by Rubin and the DDF cadence.  To model a sawtooth, we used the reverse of the \texttt{Scipy.signal} sawtooth wave, such that there is a ``burst'', or sudden increase, in brightness, and then a gentle decrease. An example of the sawtooth signal can be seen in the second panel of Figure \ref{fig:LCassembly}.

We allow the light curves to have a combination of both signals through the ``sawtoothiness" parameter of Table \ref{tab:paramgrid}. If ``sawtoothiness" is 1.0, the binary signal is a pure sawtooth signal. If it was 0.0, it is a smooth sine. Anything in between is a convolved signal of the two models that takes the form:
    \begin{equation}
    \begin{aligned}
        \Delta m(t) = (1-S) * A \sin\left( f t + \phi\right) \\
        +~ S * A\ \textrm{sawtooth} \left( f t + \phi\right) ,
    \end{aligned}
    \end{equation}
where $\Delta m$ is the change in magnitude, $m$, as a function of time $t$ with parameters $A$ (amplitude), $f$ (frequency, $f=2\pi/P$), $\phi$ (phase), and $S$ for the sawtooth parameter percentage. Panel two of Figure \ref{fig:LCassembly} shows examples of the convolved signal. The binary signal for each simulated light curve is drawn from a uniform distribution of sawtoothiness between 0 and 1. 

\begin{figure*}[p]
\centering
\noindent\includegraphics[width=\textwidth]{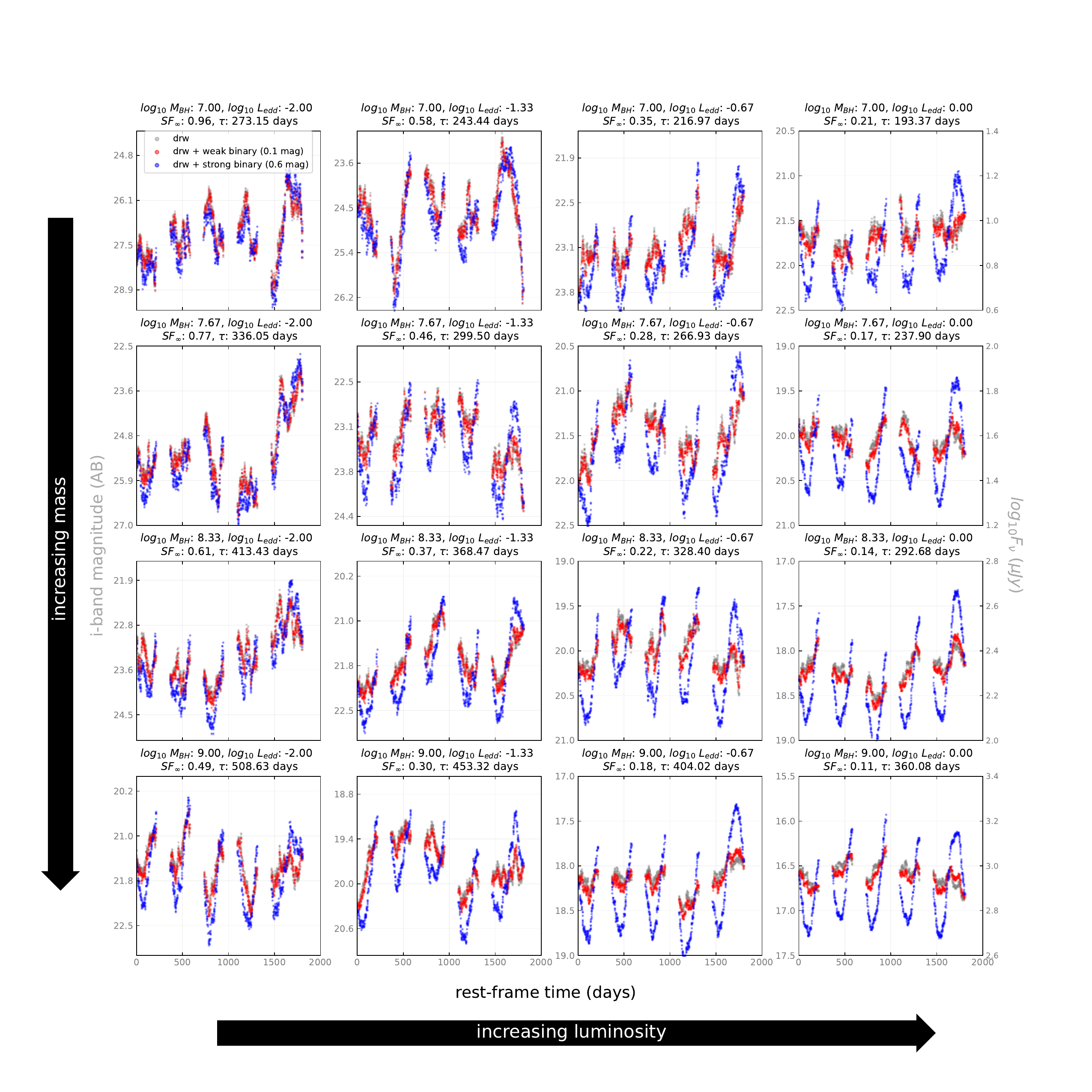}
\caption{Examples of synthetic 5-year LSST quasar light curves with seven month seasonal duration, featuring a single quasar (grey) and two potential binary systems with varying amplitudes (red: weak, blue: strong). Periodic binary signals can be difficult to distinguish from quasi-periodic quasar variability. As seen in the upper left (low mass and low luminosity quasar variability), a strong binary signal looks similar the underlying DRW. With a high mass and high luminosity quasar, seen in the lower right, variability amplitudes can be suppressed enough to directly mimic a ``weak" binary signal.  Moving down the y-axis, the black hole mass increases, moving from left to right across the x-axis, the luminosity increases. Representative values of these two properties were plugged into the DRW best-fit coefficient formulae of \cite{MacLeod_2010} and \cite{MacLeod2012} to generate the overlaid quasar variability.
Reliable binary detection with sinusoidal curve fits will be a strong function of quasar properties.} \label{fig:binaryexample}
\end{figure*}

Figure \ref{fig:LCassembly} displays full light curves in the bottom panel, in addition to the binary signals described in this section in the second panel. More examples of full light curves, with and without binaries, over a range of quasar masses and luminosities can be seen in Figure \ref{fig:binaryexample}. This figure especially highlights how much a DRW can hide or mimic a binary signal.

\subsection{ Rubin Deep Drilling Field Observational Parameters}\label{sec:RubinLSSTLCs}

The Deep Drilling Fields are the most interesting Rubin/LSST pointings for identifying binary quasars. The DDFs are five observation fields (COSMOS, XMM-LSS, ELAIS-S1, EDF-S, and CDF-S) that will receive high-cadence, multi-band observations and also already have extraordinary multi-wavelength coverage over the last two decades.
They are expected to have an average cadence\footnote{In detail, the Rubin DDF cadence is planned to cycle through each of the $ugrizY$ filters every $\sim$3 nights, with different filters observed each night \citep{brandt2018LSSTwhitepaper, Bianco_2022LSST}. The detailed DDF cadence may also follow an accordion pattern rather than a constant cadence \citep{Scolnic_2018}. We simplify by assuming that observations in different filters can be combined to create a higher-cadence light curve.} of 1-3 night(s) over the full 10-year baseline of the survey \citep{brandt2018LSSTwhitepaper, Ivezi__2019, Bianco_2022LSST}. 
Their long baselines and high cadences will provide the best opportunity for electromagnetic detection of binary SMBHs.  We can naively estimate a binary population for the LSST DDFs by extrapolating from the binary candidate population of current/previous time-domain monitoring campaigns, like that of the Catalina Real-Time Transient Survey (CRTS, \citealp{CRTS}). \cite{Graham2015} found 111 CRTS candidate objects out of 243,500 spectroscopically confirmed quasars. For the DDFs, 
this implies about 20 binary candidates, although the higher depth and signal-to-noise of the DDFs compared to CRTS means that binaries may be identified in an entirely different parameter space such that this is an uncertain prediction.

We make our simulated light curves representative of Rubin/LSST DDF monitoring by adding flux uncertainty (including the effects of weather and lunation, described further below) and masking to match the anticipated cadence and duration. We apply a seasonal monitoring duration of at least seven months out of a year to simulate light curves for one of the equatorial DDFs (COSMOS or XMM-LSS). The seasonal monitoring duration would increase for either of the fields at lower declination (ELAIS-S1 or CDF-S).



The photometric error for each light curve was calculated with Equations 4 and 5 from \cite{Ivezi__2019}. From their Equation 4, the error of a single visit, $\sigma_{1}$, is found from adding the random photometric error, $\sigma_{rand}$, and the systematic photometric error, $\sigma_{sys}$ in quadrature.
The LSST team predicts that $\sigma_{sys} < 0.005$ mag. We set it to $0.003$ mag. Then $\sigma_{rand}$ can be found from their magnitude-dependent Equation 5:

\begin{equation}\label{eqn3}
     \sigma^2_{rand} = \left ( 0.04 - \gamma \right)x + \gamma x^2 \left ( mag^2\right)
\end{equation}
where $\gamma$ is a unitless filter-dependent sky brightness parameter and $x \equiv 10^{0.4 \left( m-m_5\right) }$ with $m_5$ being the $5\sigma$ depth for a point source. 
For our $i$-band light curves, $\gamma = 0.039$ and $m_5 = 23.92$, as described in Table 2 of \cite{Ivezi__2019}. The $x$ parameter leads to sources with $i \gtrsim 23.5$ having exponentially increasing flux uncertainties. We ultimately apply a $i<23.5$ quality cut on our light curves as a representative range for LSST light curves of sufficiently high quality for binary characterization.

We make these light curves more realistic by further including error multipliers for the moon phase and environmental factors such as weather and photometric conditions. The moon phase error multiplier is a sinusoidal function based on the 28-day moon cycle, with values ranging from 1.0x (dark time, new moon) to 2.5x (bright time, full moon). The weather multiplier was a random, discrete choice between 0.5x (excellent photometric conditions), 1.0x (average), and 1.5x (poor photometric conditions), randomly applied to the light curves with weights of 0.15, 0.75, and 0.15, respectively. A weather multiplier for variations in seeing on an observation-to-observation basis, with the average resulting in a multiplier of 1x. 
We also simulate observations lost due to weather or engineering time by randomly removing 30\% of the points in each light curve.

Once the variability model and binary signal were added together, the individual points were then randomly resampled within their photometric error and replaced (bootstrap re-sampling). The bootstrap re-sampling was done by randomly selecting a new magnitude from a normal distribution of a single observation's magnitude and its corresponding error. 

Summarizing, we apply the following treatment to the light curves to resemble anticipated LSST DDF light curves:
\begin{itemize}
    \item 1-day cadence
    \item 7-month seasonal duration
    \item Flux uncertainty predicted for LSST \citep{Ivezi__2019}
    \item Additional flux uncertainty associated with lunation and (random) seeing conditions
    \item 30\% (random) cadence losses due to weather
\end{itemize}

This final stage of the light curve generation could be easily adapted to the cadence, duration, and noise properties of other surveys. In this work we focus on the upcoming LSST DDF monitoring, as the most efficient near-future survey for discovering binary quasars, but in future work we anticipate applying our pipeline to both the LSST main survey and the lower cadence and 
noisier monitoring observations of CRTS. If we assume a similar number of binary candidates as identified in \cite{Graham2015}, then the LSST WFD main survey could have approximately 23,000 candidate binaries which is promising for a photometric binary detection confirmation in the near future.


\section{Light Curve Fitting}\label{sect:Analysis}

We chose to implement simple sinusoidal curve fits to the light curves as the detection method in this work. This method is computationally inexpensive and can provide insight into both binary parameter recovery rates and false positive rates. A sine function fit is a common method used in finding binary SMBHs \citep{Graham2015, Witt_2022}.
The goal of curve fitting the data was to characterize the ability to recover the amplitude, phase, and period of the binary signal from the simulated DRW + binary light curves sampled with the expected observation pattern for DDFs. We fit the same sine function to the isolated-quasar (pure-DRW) light curves to estimate the false positive detection rate of periodic binary variability. 
Our computationally inexpensive method could be used as a first ``filter" to triage data before applying multi-component and computationally expensive methods, such as the Bayesian model selection analysis of \cite{Witt_2022}. 

\begin{figure*}[t]
\centering
\noindent\includegraphics[width=\textwidth]{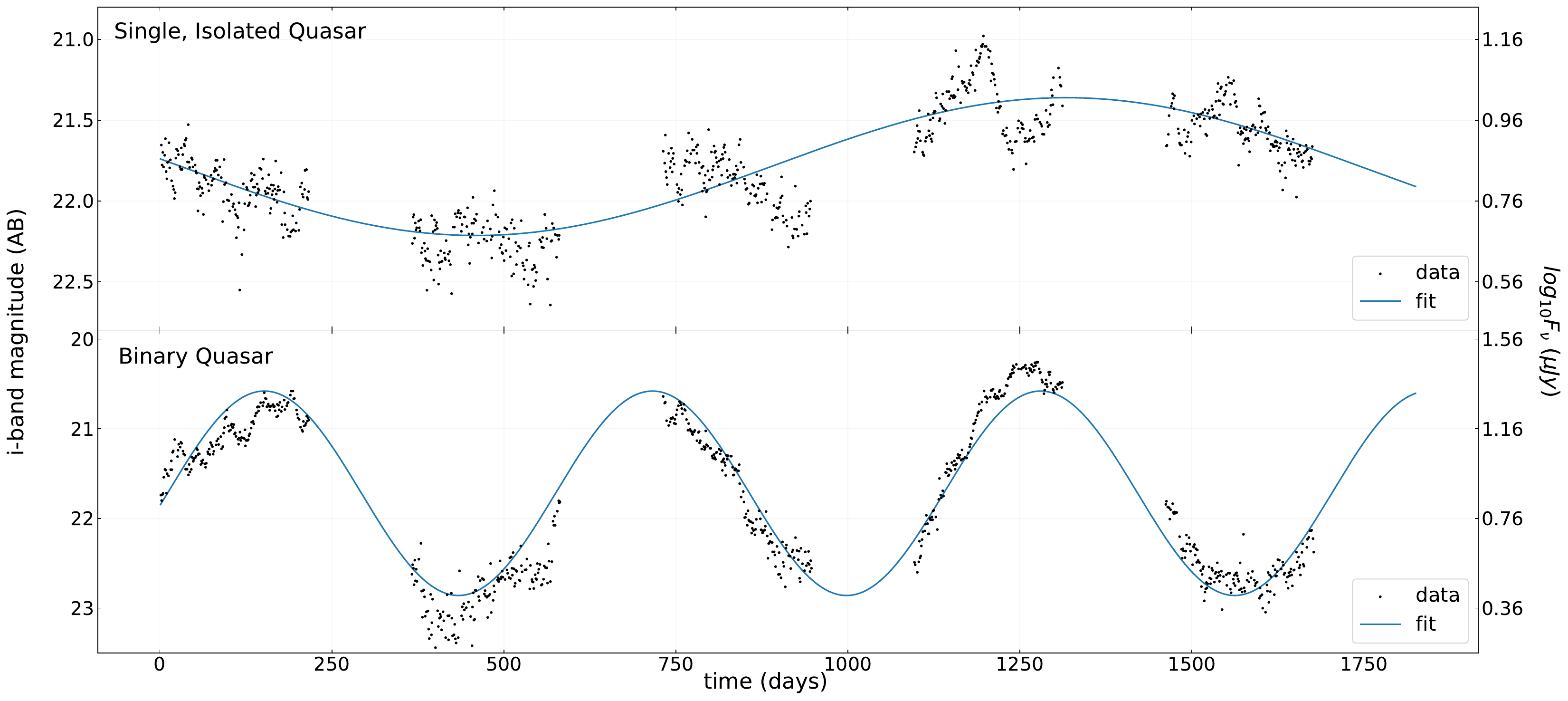}
\caption{Examples of simulated LSST Deep Drilling Field light curves, with and without binaries, and their respective fits. Both light curves have underlying variability associated with a $10^8$ $M_{\odot}$ quasar at z $\sim$ 1 that is accreting at 10\% of its Eddington limit. \textit{Top:} The best sinusoidal fit to a non-binary, isolated quasar. \textit{Bottom:} The best sinusoidal fit to a binary quasar with a signal amplitude of 0.9 mag, period of 570 days, and a phase of $\pi$. }\label{fig:ExampleFits}
\end{figure*}

We utilized the Python package \texttt{LMFIT}, a non-linear least-squares minimization and curve-fitting tool for Python that utilizes the Levenberg–Marquardt algorithm, for the curve fits \citep{lmfits}. Examples of these fits to simulated light curves can be seen in Figure \ref{fig:ExampleFits}.
We used their built-in \texttt{SineModel} to fit magnitude $m$ as a function of time $t$ with parameters $A$ (amplitude), $f$ (frequency, $f=2\pi/P$), and $\phi$ (phase):
    $m(t) = A \sin\left( f t + \phi\right)$.
We do not fit a sawtooth function to the light curves, only the sinusoid. This is to investigate how much confusion missing the sawtooth might introduce.

\begin{figure*}[t]
\centering
\noindent\includegraphics[width=\textwidth]{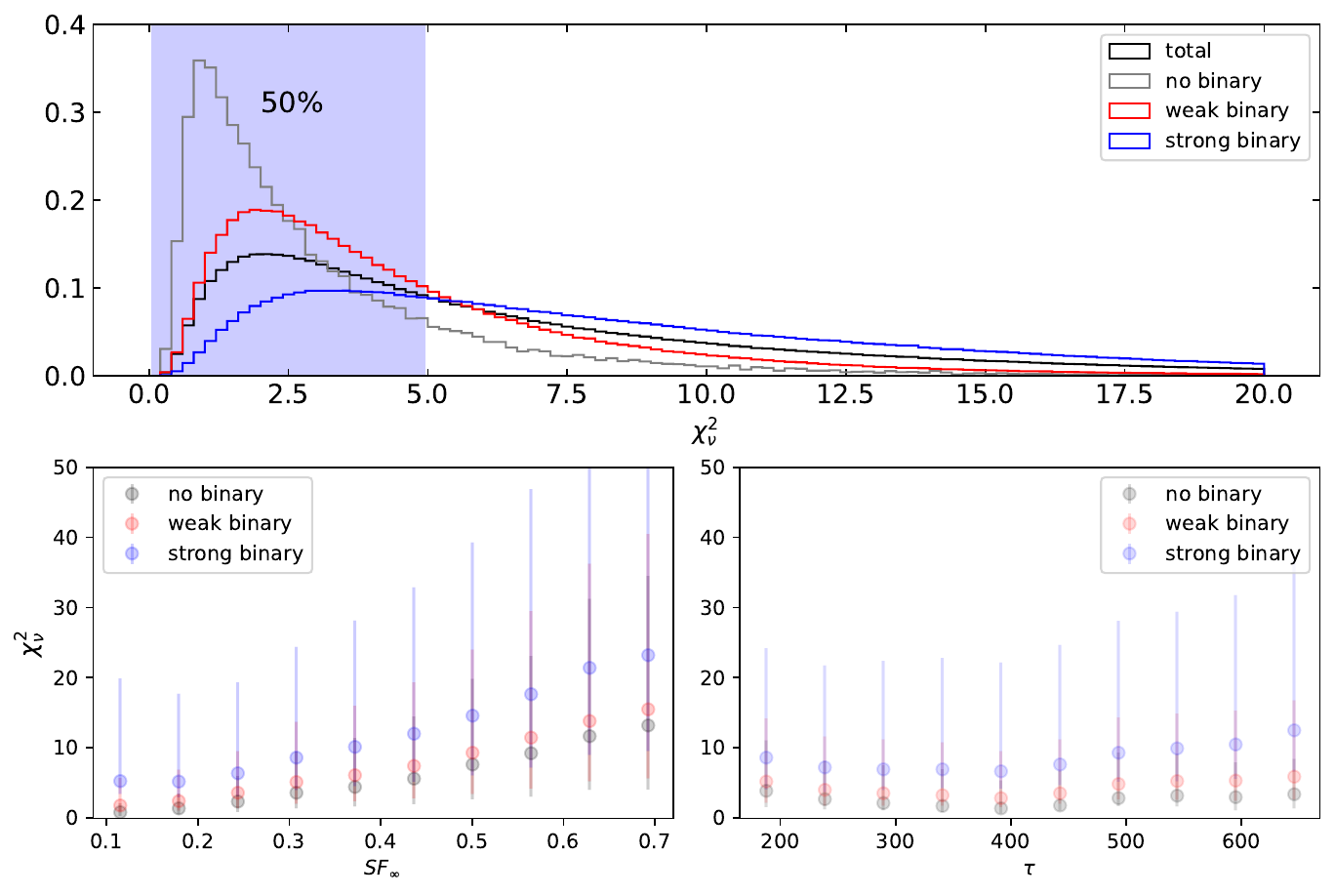}
\caption{\textit{Top:} The normalized distribution of reduced chi-squared for the sine fits to the simulated light curves, separated by light curve type. Shaded region indicates that 50\% of the total distribution has a $\chi_\nu^2 \lesssim 5.0$. \textit{Bottom:} The distributions of reduced chi-squared with the DRW amplitude (left) and damping timescale (right). Plotted are the median y-axis value for the data binned along the x-axis. The error bars shown are the 16th and 84th percentile values for the binned data. The fits are worse when the DRW amplitude is strong, but there is not a simple monotonic relationship with $\tau$. The best-fit sine function is generally a good fit to the data, especially for light curves with strong binary and/or weak DRW contributions. We note that the fits for strong binaries are generally ``worse" in terms of \redchi, this is a consequence of large values in the sawtoothiness parameter.}\label{fig:LCchi}
\end{figure*}

We evaluate goodness-of-fit using the reduced chi-squared statistic \redchi.
We introduced an error floor of 0.5\% of the light curve's mean magnitude, to be used as a fit weight, for the reduced chi-squared evaluation to avoid unreasonable \redchi values for bright quasars (that have very small uncertainties from Equation \ref{eqn3}). 
The relative \redchi ranges and visual inspection results can be seen in Figure \ref{fig:LCchi}.   
Fitting a sine to a DRW-only light curve gives a better numerical fit than an DRW+sine because of the multi-component interaction of the two models (i.e., different damping timescale and binary period values produce light curves that are not well-fit by a single sine function). Larger amplitude binary signals, especially sawtooth signals, worsen \redchi values overall as well.

For the following analysis, we define a ``weak'' binary as one with an injected binary signal amplitude $0.1 < A \leq 0.5$~mag and a ``strong'' binary with amplitude $0.5 < A < 1.0$~mag. We note that the ``weak" label reflects binary signals that are actually quite moderate when considering most candidate binaries have amplitudes near 0.1~mag \citep{Graham2015}. This label is only introduced for ease of the following analysis. The bottom panels of Figure \ref{fig:LCchi} demonstrate that the sine functions provide better fits (lower $\chi_\nu^2$) to the simulated light curves with weaker DRW amplitudes and stronger binary signals as expected. The sine function fits are also better when the DRW damping timescale is longer that the underlying binary period, likely because it makes the shorter binary period easier to distinguish from the quasi-periodic DRW variability.

The next section discusses how the rates of binary detections and false positives depend on quasar and binary properties.
We emphasize that these simple sinusoidal fits are computationally inexpensive: we achieved $\sim$7,500 light curve fits per minute on a modest 4-core personal laptop. This makes the fits well-suited to survey data applications like the LSST event broker\footnote{https://www.lsst.org/scientists/alert-brokers (\today)}. 

\section{Results}\label{sec: results}
\subsection{Binary Signal Recovery}

\begin{figure*}[t]
\centering
\noindent\includegraphics[width=\textwidth]{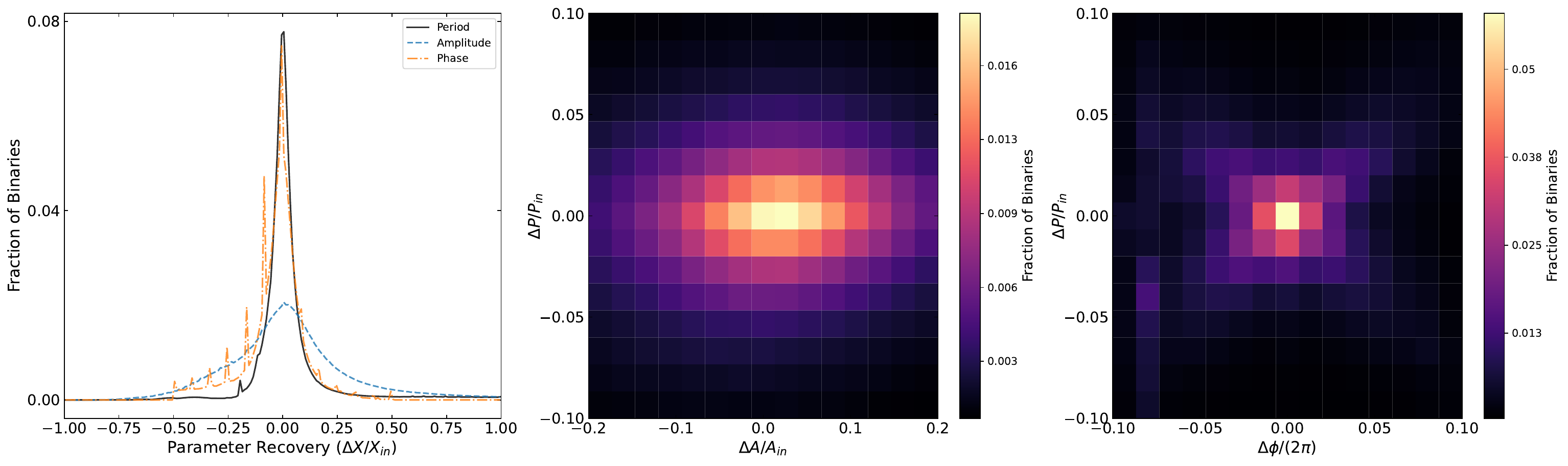}
\caption{\textit{Left:} Fractional binary parameter recovery performance for amplitude ($A$), period ($P$), and phase ($\phi$) for all binary types. For all binaries, 28\% can have all three parameters recovered within 10\% (0.10) of their input values. The phase spikes are a consequence of possible negative amplitudes when fitting and phase folding.  \textit{Middle:} 2-D histogram of the period and amplitude recovery to show any parameter dependence. Period and amplitude recovery performance are generally uncorrelated. Only 30\% of all binaries can have both the binary signal amplitude and period recovered together. \textit{Right:} Period and phase recovery dependence. Where period is likely recovered, the phase is recovered, especially within 10\% of their input values. Over 61\% of all binaries can have their signal phases and  periods recovered within 10\%. }\label{fig:ParamRecover}
\end{figure*}

We first investigate the ability of our sinusoidal fits to recover the input binary signal. For most parameters we characterize the accuracy of recovery as a fractional error: (fit parameter output - parameter input) / (parameter input), i.e. $\Delta{P}/{P_{in}}$. Phase recovery is instead quantified by the fraction of a period, $\Delta\phi/(2\pi)$.
Figure \ref{fig:ParamRecover} shows the fit recovery of the input binary period ($P$), amplitude ($A$), and phase ($\phi$). The left panel shows the distribution of individual parameter recovery: 
binary period and phase have similar recovery rates,   
while amplitude is more poorly recovered.  The right panels of Figure \ref{fig:ParamRecover} demonstrate that the phase and period are recovered in tandem: where the period is generally well-recovered, the phase is as well. The accuracy of amplitude recovery is independent of the period and phase recovery. This
is due to the additional variability of the DRW. 

\begin{figure*}[t]
\centering
\noindent\includegraphics[width=\textwidth]{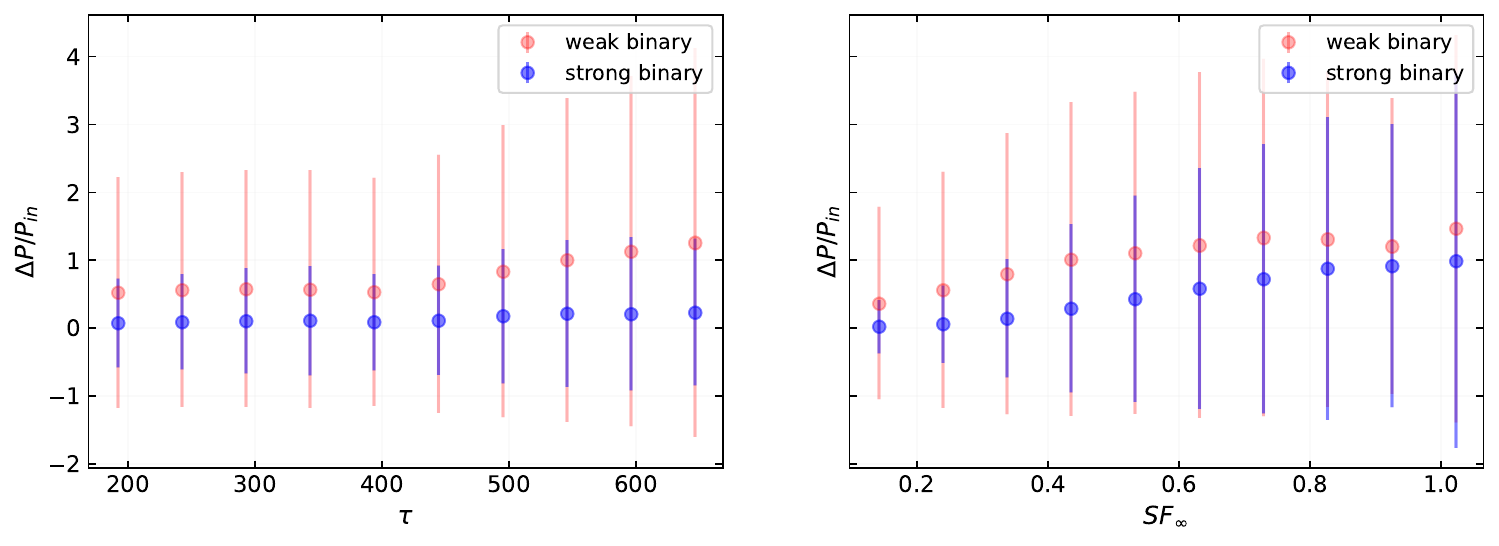}
\caption{ Period recovery, binned by DRW parameters (\textit{left:} damping timescale $\tau$, \textit{right:} DRW amplitude \SFinf). Plotted here, and in all similar plots that follow, are the mean y-axis values for the data binned along the x-axis. The error bars shown are
the standard deviations within those bins. The period is overestimated for weaker binaries (by factor of 1.5-2) because the red noise variability acts to smooth out the periodic signal, leading to best-fit sinusoids with longer periods. This is seen in both parameters. The stronger the DRW amplitude, the more overestimated the best-fit periods become, regardless of binary type.}\label{fig:periodVdrwparams}
\end{figure*}

Period is generally a well-recovered parameter, with 71\% of the fits to all binary light curves recovering the input period with $\Delta{P}/P_{\rm in} <$ 0.10 (or 10\%). Binary signals with strong (0.5 $< A <$ 1.0) amplitudes can have their period recovered 84\% of the time. Figure \ref{fig:periodVdrwparams} shows that binary period is modestly overestimated by the fits for quasars with larger DRW amplitudes. It also shows that all binary types are most affected by the DRW amplitude, rather than the damping timescale.  

\begin{figure*}[t]
\centering
\noindent\includegraphics[width=\textwidth]{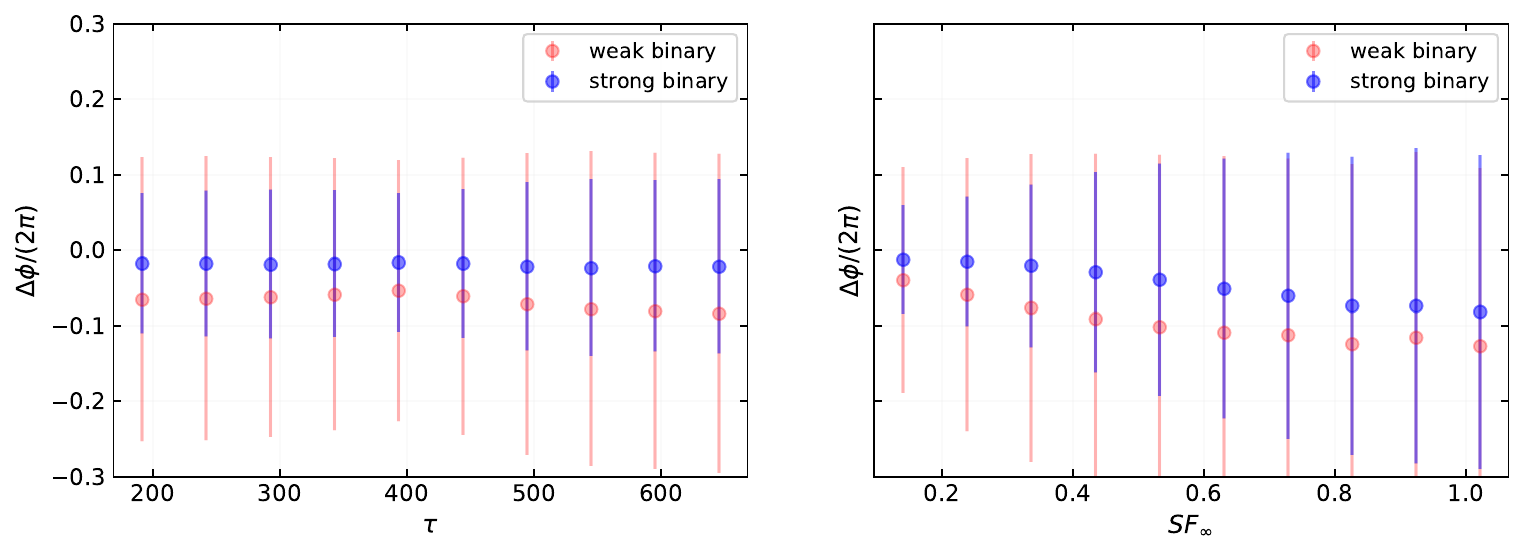}
\caption{Fractional phase recovery means (points) and standard deviations (error bars) as binned by DRW parameter (\textit{left}: damping timescale $\tau$, \textit{right}: DRW amplitude \SFinf). Phase recovery is modestly underestimated across both DRW parameters. It has the strongest dependence on \SFinf, for all types of binaries. } \label{fig:phaseVdrwparams}
\end{figure*}

The binary phase is recovered as well as the period, with a trend of underestimation. Roughly 69\% of all binary light curves had their phase recovered $\Delta{\phi}/(2 \pi) < $ 0.10, with strong binary phase recovery at 81\%. Figure \ref{fig:phaseVdrwparams} shows that again the DRW amplitude has more influence on phase recovery than the damping timescale. While period and phase recovery are generally correlated, the phase is less affected by DRW parameters than the period for larger values of \SFinf\ and $\tau$.  This is evident in the lack of strong correlations and small standard deviations of $\Delta \phi / (2 \pi)$ with DRW parameters.

\begin{figure*}[t]
\centering
\noindent\includegraphics[width=\textwidth]{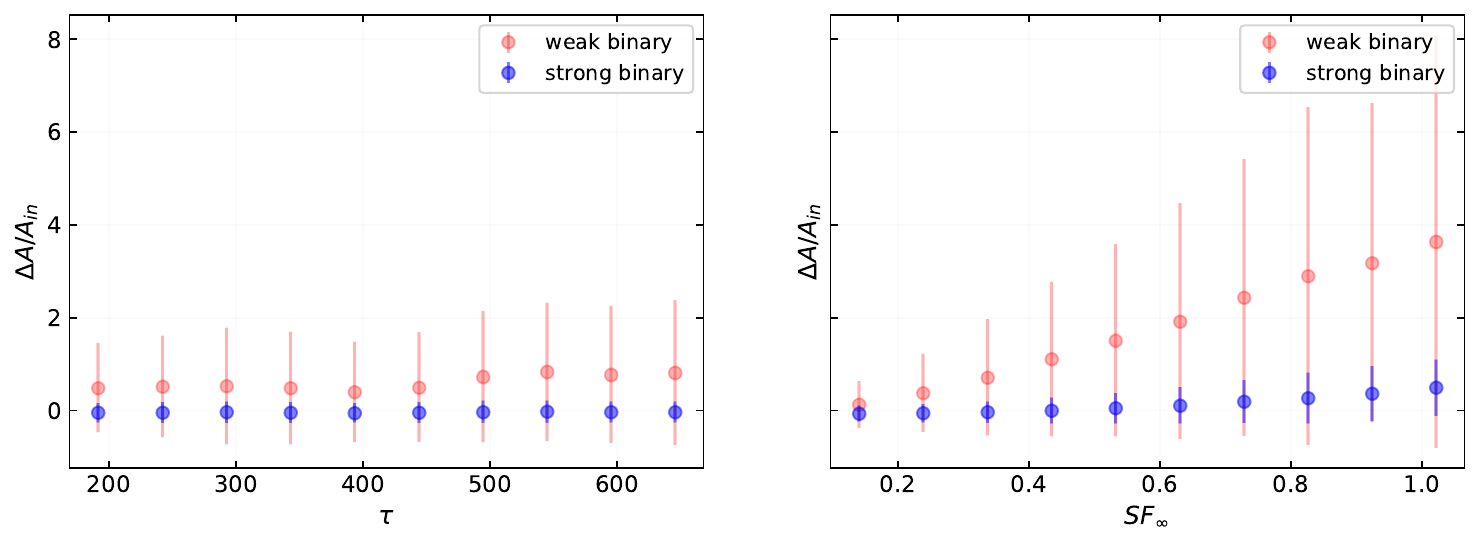}
\caption{Fractional recovery of binary amplitude, binned by the DRW parameters (\textit{left:} damping timescale $\tau$, \textit{right:} DRW amplitude \SFinf). The binary amplitude is effectively recovered for light curves with strong binary signals (binary amplitude $A>0.5$~mag), with little dependence on the DRW signal. Light curves with weak binary signals generally have best-fit amplitudes that are significantly overestimated, especially for large DRW amplitude.} \label{fig:ampVDRWparams}
\end{figure*}

Binary amplitude is the least effectively recovered parameter, with only 36\% of the fits recovering $\Delta{A}/A_{\rm in} <$ 0.10. Strong binaries see a higher recovery rate at 44\%.
Figure \ref{fig:ampVDRWparams} shows the binary amplitude recovery when binned by DRW parameters. Amplitude recovery has the same general trends as phase and period, with the DRW amplitude having the strongest effect on recovery. The binary amplitudes are generally overestimated when the DRW amplitude is larger. The weak binary behavior shows that as the DRW amplitude becomes many more times the binary signal amplitude, the binary signal becomes more difficult to recover. Only 28\% of all binary light curves can have all three binary signal parameters recovered at the same time to within 10\% of their input values.
When the binary amplitude is near or larger than the DRW amplitude (e.g., strong binaries with $A>0.5$ and \SFinf~$<0.5$), recovery for all parameters is better, as similarly pointed out by \citet{Witt_2022}.
In general the period and phase can be recovered even when the amplitude cannot.  

\begin{figure}[t]
\centering
\noindent\includegraphics[width=\linewidth]{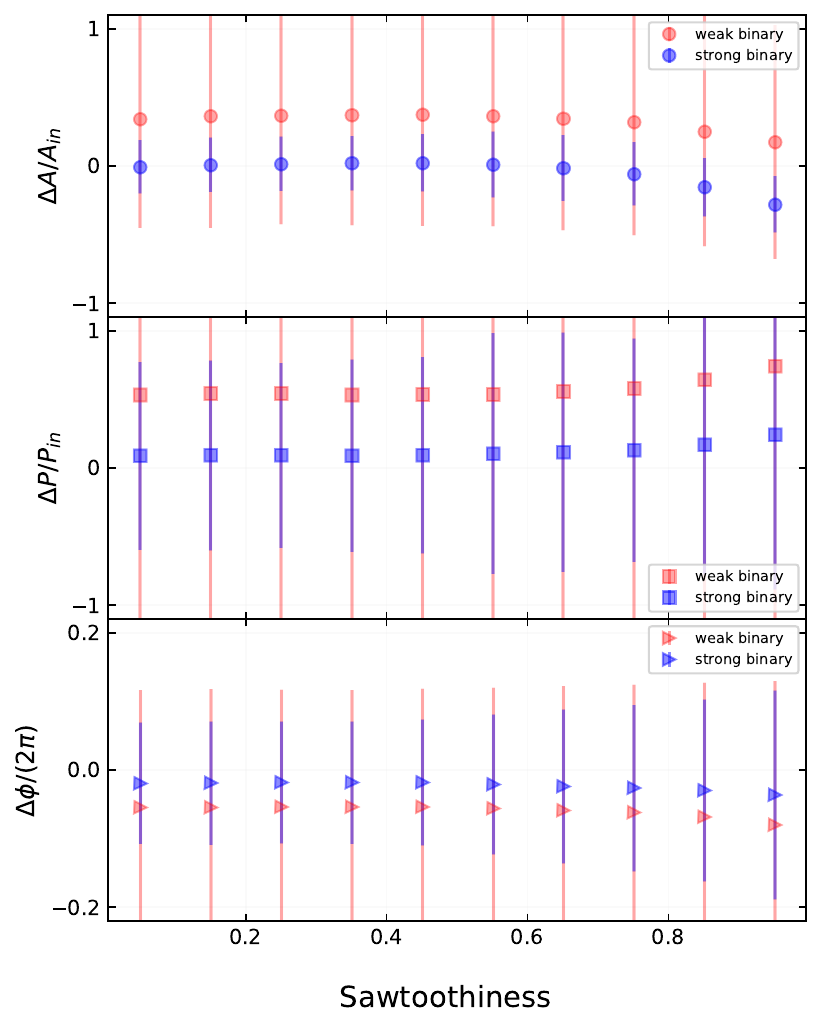}
\caption{Parameter recovery compared to binary signal ``sawtoothiness'' (0.0 is a smooth sine, 1.0 is a pure sawtooth).
From top to bottom, panels show the parameter recovery for binary amplitude $A$, period $P$, and phase $\Phi$ for weak (red points) and strong (blue) binaries. In all cases, the stronger the binary signal, the better the recovery is.  There is interesting behavior at large values of sawtoothiness: the period and amplitude become more over- and under- estimated, respectively, by 10-20 \% compared to their smoother sine counterparts. Despite this, DRW parameters have a stronger influence on recovery than signal type/structure.} \label{fig:RecoveryVSSawtooth}
\end{figure}

Next, we investigate the effectiveness of the binary signal parameter recovery as a function of binary signal structure. Figure \ref{fig:RecoveryVSSawtooth} depicts the recovery performance of each parameter as a function of sawtoothiness. The sinusoidal fits
still accurately recover the parameters even if the 
light curve is quite ``sawtoothy.'' As discussed above with the fit results versus DRW parameters, the amplitude and period are again overestimated and phase is underestimated, with weaker binaries displaying larger deviations for all three parameters. Larger values of sawtoothiness drive the best-fit values of amplitude and phase lower compared to their smoother sine counterparts.
However, these differences are small at 10-20\%. The DRW parameters have a stronger correlation with binary parameter recovery than the binary signal structure type.

\subsection{Non-Binary Quasar Fit Results}\label{sec:Nonbins}

\begin{figure*}[t]
\centering
\noindent\includegraphics[width=\textwidth]{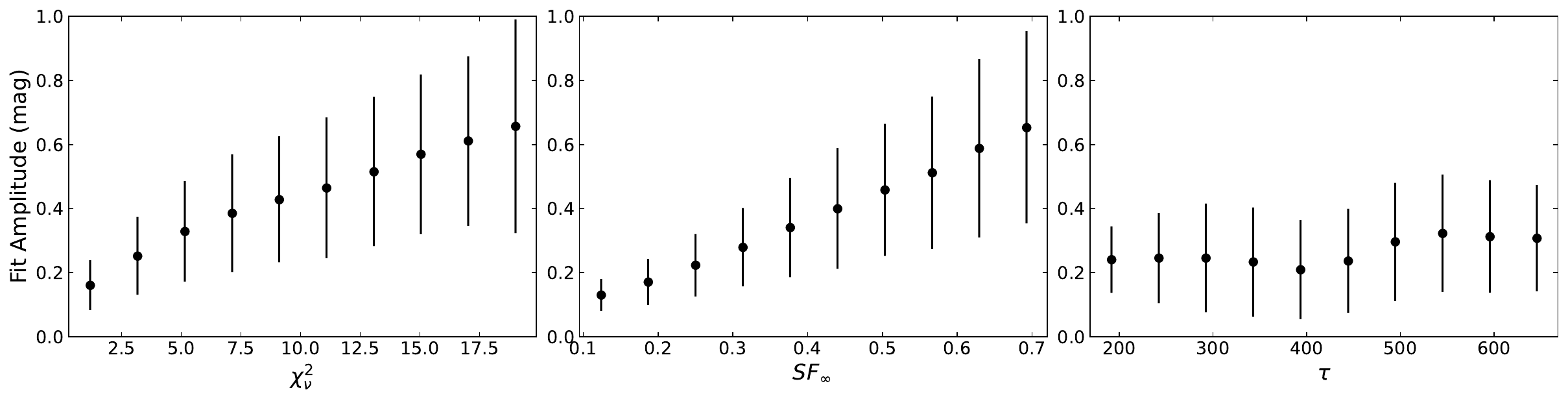}
\caption{Best-fit amplitude results for single, isolated quasars.  \textit{Left:} Single quasars that are well-fit by the sinusoidal binary light curve model (with low \redchi) generally have small best-fit amplitudes.
\textit{Middle:} Best-fit amplitude and \SFinf have a positive, linear correlation: the stronger the DRW amplitude, the larger the sinusoidal fit amplitude. \textit{Right:} The is no significant connection between fit amplitude and damping timescale, $\tau$. }\label{fig:NonbinAmp}
\end{figure*}

We next investigate how the single, isolated quasars performed when fit with the periodic functions. Figure \ref{fig:LCchi} shows that more than 80\% of the single quasar light curves have a best-fit model with \redchi $< 5.0$. 
We further explore what the best-fit results are for these light curves.
Figure \ref{fig:NonbinAmp} depicts the best-fit amplitude versus \redchi, DRW amplitude, and damping timescale for the single (non-binary) quasar light curves. The fit amplitudes directly correlate to DRW amplitude, with slight underestimation. The sinusoidal fits have a difficult time fitting larger \SFinf, as most light curves with a \redchi $\leq$ 5.0 have best-fit amplitudes  0.1 $< \rm A <$ 0.6. From Figure \ref{fig:binaryexample} and the results of the previous section, most of the quasars with large structure functions would obscure a binary signal and make recovery of the underlying parameters difficult. 

\begin{figure*}[t]
\centering
\noindent\includegraphics[width=\textwidth]{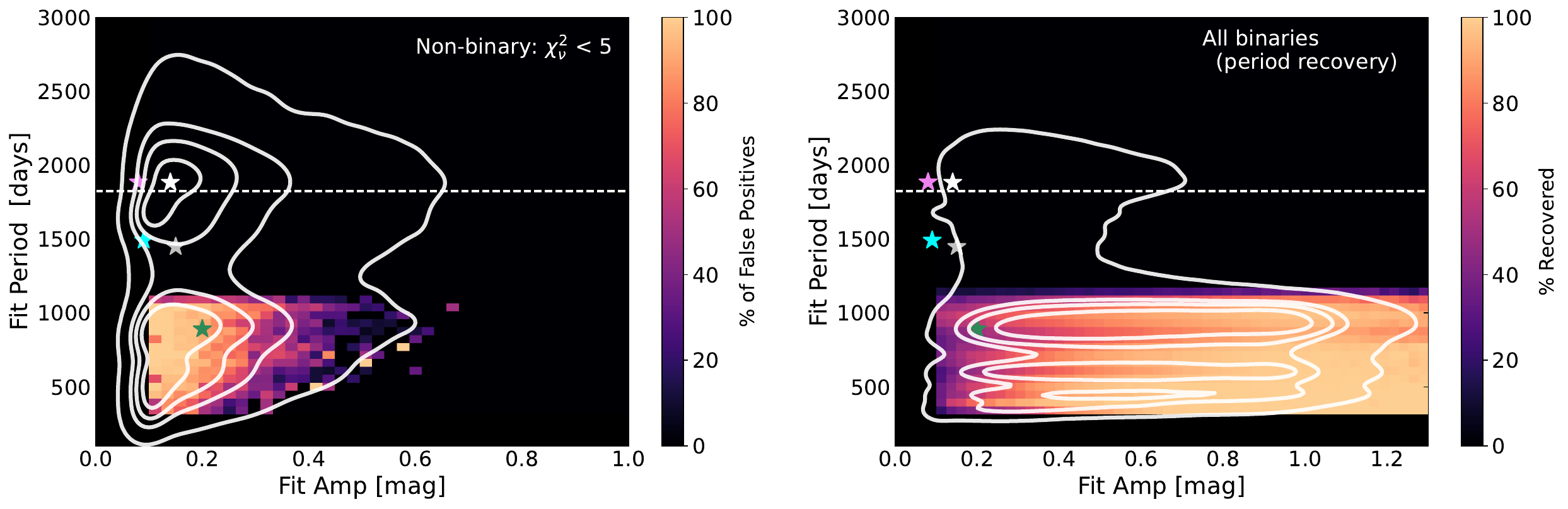}
\caption{Best-fit amplitudes and periods of sinusoidal binary model fits to non-binary quasars (\textit{left panel}), shown in the contours with false positive rates shown in the underlying 2-D histogram. The best-fit amplitudes are small and the periods have a bimodal distribution with peaks around 900 and 1800 days with preference for a longer period as a consequence of the length of our light curves. The upper bound on the false positive rate is also tied to light curve and survey duration. \textit{Right:} the best-fit amplitudes and periods of our binary light curves shown in the contours, with recovery rates (using period recovery) of the full binary population shown in 2-D histogram.
Section \ref{sec:FPs} defines our false positive threshold. The white dashed lines in both plots indicate the duration of our light curves: five years. The stars are example binary candidates, discussed in Section \ref{sec:RubinRecs}.}\label{fig:NonbinDemo}
\end{figure*}

Amplitude has a clear dependence with DRW parameters, as seen from Figure \ref{fig:NonbinAmp}, and while best-fit period has no relationship with the DRW parameters, it is a non-uniform distribution (see contours of Figure \ref{fig:NonbinDemo}). Phase was not investigated further as its distribution was uniform and showed no dependence across \redchi, \SFinf, and $\tau$. 

Investigating the best-fit binary models for non-binary quasars is important because single quasars are likely to vastly outnumber binary quasars \citep{Kelley2019c}. If binary quasars are selected to have $\rm A >$ 0.1, as reflected by the candidate populations in previous surveys (e.g. \citealt{Graham2015}), then putative samples of binary quasars will include (or be dominated by) a large number of false positives.
More discussion of false positive rates and population demographics can be found in Section \ref{sec:FPs}.

\section{Discussion} \label{sec: discuss}

In the following section, we discuss the general trends seen in the data. We further define what kinds of binary SMBH detections can be trusted and introduce false positive detection rates, provide recommendations for binary SMBH searches from Rubin/LSST light curves, and discuss the multi-messenger connections of these results.

\subsection{Implications for True Binary Recovery}

Our analysis of binary parameter recovery establishes that fitting simple (and computationally inexpensive) sinusoidal binary light curve models usually recovers the period and phase to within 10\% of the input value (Figure \ref{fig:ParamRecover}). We recover the period from 84\% of our strong binary light curves and 71\% of all binary light curves.  Figure \ref{fig:periodVdrwparams} establishes that smaller values of damping timescale and DRW amplitude make it easier to accurately recover the binary signal.
The phase can be recovered for 81\% of all strong binaries and 69\% across all binaries. Phase is typically underestimated and most affected by DRW amplitude. Binary signal amplitude is not effectively recovered for most light curves, as only 44\% of strong binaries and 36\% of all binaries can have their amplitudes accurately recovered within 10\% of the input value. Both phase and period can be recovered simultaneously for 62\% of all binaries. Sinusoidal curve fits can only recover all three binary parameters for 28\% of all binary light curves.

In general, the ``stronger'' (higher amplitude) the binary signal, the easier it is to disentangle from a DRW. This remains true even if the DRW amplitude is large. This is aligned with the results of \cite{Witt_2022}. Weaker binaries are more affected by strong DRW parameters and best-fit amplitude especially is overestimated because of the additional variability of a strong DRW signal. The true binary amplitude will be smaller than measured for most binary quasars by about a factor of 1.5-2 because of the contribution of DRW variability. While strong binaries often have their binary signal parameters recovered within 10\% of the true value, we note that most binary candidates in the literature (e.g., \citealp{Graham2015, Charisi2016}) have amplitudes that fall within our ``weak'' binary category.

Simple, computationally inexpensive fits are ambivalent to sawtoothiness for a wide range of binary signal parameters, as in the binary parameters can still be recovered. The contribution from the DRW matters more than the details of the binary signal. This is encouraging news as we do not know the true nature of these binary signals (see Section \ref{sec:binarysig} for description of the different types and emission mechanisms of electromagnetic binary signals). The accuracy of period and phase recovery is promising as they can be directly linked to the gravitational wave signal parameters \citep{Charisi_2022}. Section \ref{sec:MMA} offers more discussion on joint EM and GW endeavours.

\subsection{False Positive Detection and Reliability Thresholds}\label{sec:FPs}
\begin{figure*}[t]
\centering
\noindent\includegraphics[width=\textwidth]{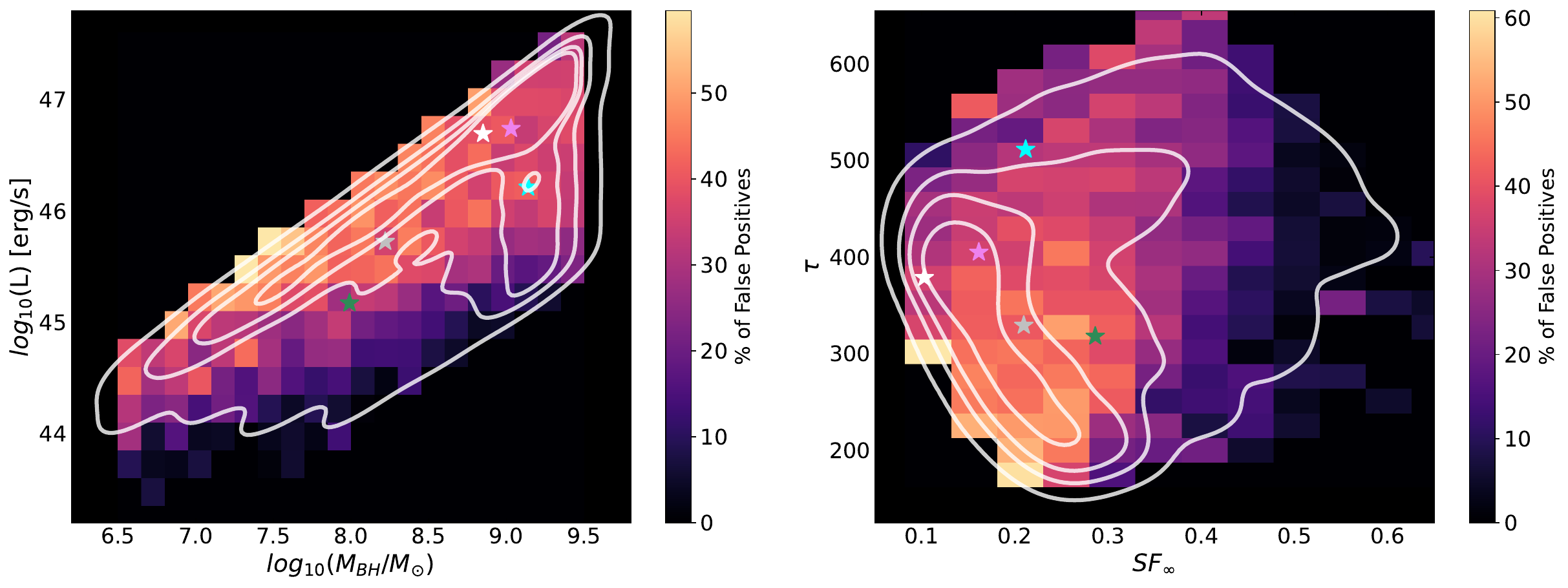}
\caption{2-D histograms of the false-positive demographics, contours (white) of the full non-binary population, and stars of five example binary candidates from Table \ref{tab:Candidates}. \textit{Left:} The luminosity and mass grid of false positives. We find that the most luminous and massive quasars have the highest rates of false positive detection. \textit{Right:} False-positive rates in terms of DRW parameters, damping timescale $\tau$ and DRW amplitude \SFinf.  We having that small $\tau$ and \SFinf drive the false positive rate, with the light curves made of those values reporting upwards of 60\% false positive rates.}\label{fig:FPDemographics}
\end{figure*}

\begin{figure*}
\centering
\noindent\includegraphics[width=\textwidth]{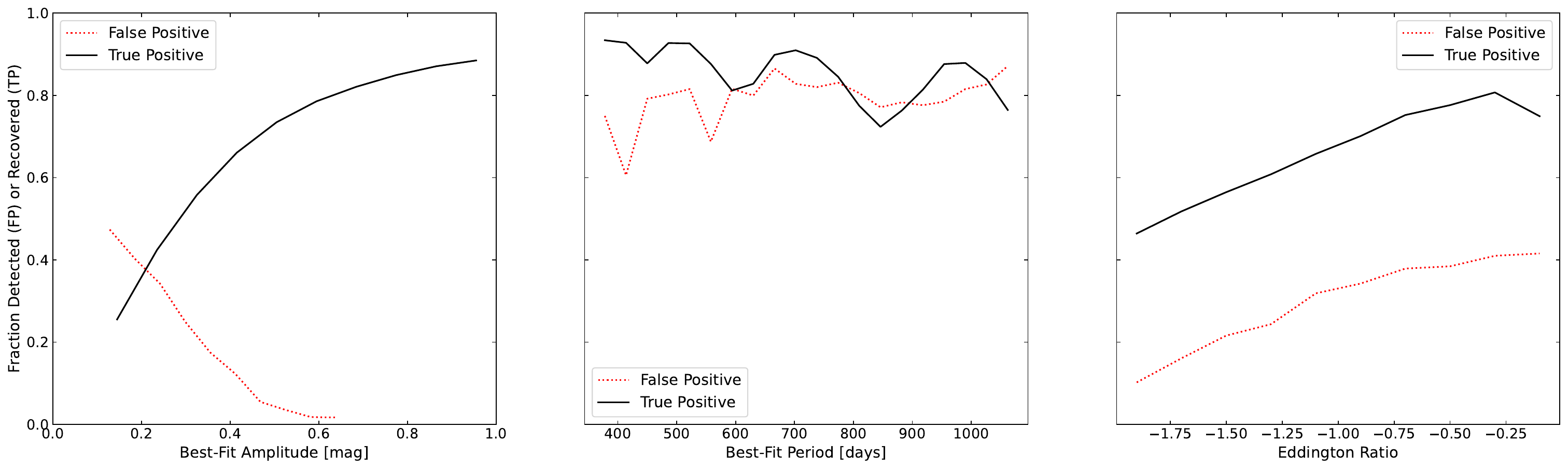}
\caption{Reliability thresholds: the best-fit amplitudes and periods of false positive binaries and the ``recovered'' fraction of true positive binaries (recovered, here, is defined as having the binary signal periods recovered within 10\% of their input value via the sinusoidal fits). Also plotted are the Eddington ratio distributions for false positives and recovered true positives. At small best-fit amplitudes, the false positive detections dominate. The stronger the binary signal, the more likely we obtain a true positive.
False positive rates are highest for quasars with high Eddington ratio.
}\label{fig:TrustAmp}
\end{figure*}

We next address the rate of false-positive detections from the simulated single-quasar (non-binary) light curves. For this purpose we define a binary detection as:
\begin{itemize}
    \item good fit by the sinusoidal light curve model, \redchi$ < 5.0$
    \item best-fit binary amplitude $A > 0.1$~mag
    \item best-fit binary period $180 < P < 1100$~days
\end{itemize}
The $A > 0.1$ requirement is motivated by the typical amplitudes of binary SMBH candidates identified from previous CRTS observations by \citet{Graham2015}. The typical amplitudes of a binary candidate population in a survey are driven by the photometric error associated with the survey. Compared to CRTS, LSST will have lower photometric error and, thus, lower resolvable binary amplitudes, but we leave this threshold as the requirement to be conservative.  The \redchi$ < 5.0$ requirement indicates that the light curve is well-fit by a sinusoidal model (this requirement is analogous to the auto-correlation tests of \citealt{Graham2015}). The range of $180 < P < 1100$ days requires that the period is well-detected (1.5 period cycles or more) within 5 years of LSST monitoring.

The left panel of Figure \ref{fig:NonbinDemo} presents the distribution of best-fit period and amplitude for the sinusoidal fits to the non-binary quasars. For comparison, the right panel of Figure \ref{fig:NonbinDemo} presents the period and amplitude parameter space for reliable recovery (defined as $| \Delta P / P | <$ 0.10) of the simulated true-binary light curves. Many of the non-binary light curves are best fit by periods of $\sim$5~yr, comparable to the survey duration and rejected by our period requirement for a binary detection. This is in contrast to the shorter periods, by construction, of the simulated true binary light curves. Most of the non-binary light curves are best fit by low-amplitude sinusoid models, again in contrast to the true binary population having most accurate recovery at high amplitude. The false-positive rate is very high for low to moderate amplitudes.
In other words, if binary candidates are selected to have well-fit sinusoid models with amplitudes of $A > 0.1$~mag, then false-positive non-binary quasars will dominate the sample.

Figure \ref{fig:FPDemographics} presents the false-positive rate as a function of quasar luminosity and mass (left panel) and DRW parameters (right panel).
The false-positive rate is highest for luminous quasars accreting near their Eddington Limit, which tend to have smaller DRW amplitudes and modestly variable light curves that can be well-fit by periodic signals.
Between 30 and 40\% of quasars with $\log(M_{\rm BH}/M_\odot)>$ 9 and $\log(L/(erg/s))>$ 47 are identified as false-positive binaries according to the selection criteria described above. For luminous quasars near a mass of  $\log(M_{\rm BH}/M_\odot) \sim$ 7.5, the false positive rate is over 50\%.
\citet{Witt_2022} demonstrate that more computationally intensive approaches can better recover binary parameters by disentangling the DRW contribution, although that work still finds a dependence of binary recovery on DRW amplitude.
 
Figure \ref{fig:TrustAmp} compares the fraction of false-positive detections (black lines) with the recovery fraction of true-positive binaries (dotted red lines), with binary ``recovery'' defined as $|\Delta P/P| <$ 10\% as before. In Figure \ref{fig:TrustAmp}, we see that at best-fit amplitudes below 0.3 mag, the false positive binaries values dominate over the true positive binaries. The higher the amplitude fit, the more likely it is that the binary is real. Single, isolated quasars have a difficult time producing a fitted amplitude above 0.4 mag and a decent \redchi statistic. We establish this as the reliability threshold for trusting a binary quasar detection via sine fits, where we can begin to trust strong binary signal amplitudes greater than 0.4~mag (see Section \ref{sec:MMA} for details of physical binary parameters for this value). We also note that a sine fit to quasar light curves will always return a non-zero amplitude value with the small predicted photometric errors of LSST.
Fits with smaller periods are more likely to be true-positive binaries, as shown in the middle panel and also because almost half of the non-binary light curves are best fit by periods comparable to the survey duration (left panel of Figure \ref{fig:NonbinDemo}). Quasars with lower Eddington ratio are also less likely to identified as false-positive binaries, with a factor of $\sim$2 decrease in the false-positive rate from \Eddratio\ $\sim$0.1 to $\sim$0.01. In contrast the rate of true-positive recovery decreases more modestly with Eddington ratio, declining by a factor of only $\sim$1.5 over the same range.

Our overall conclusions are that false positives will dominate samples of binary candidates found via sinusoidal curve fits for best-fit light curves of small amplitude ($0.1<A<0.2$~mag), large period (especially if near the survey length), and for luminous and massive quasars (with smaller stochastic accretion-disk variability). Binary candidates with larger amplitude ($A>0.5$~mag), smaller period, and lower Eddington ratio are more likely to be genuine binary quasars.

\begin{figure*}[t]
\centering
\noindent\includegraphics[width=\textwidth]{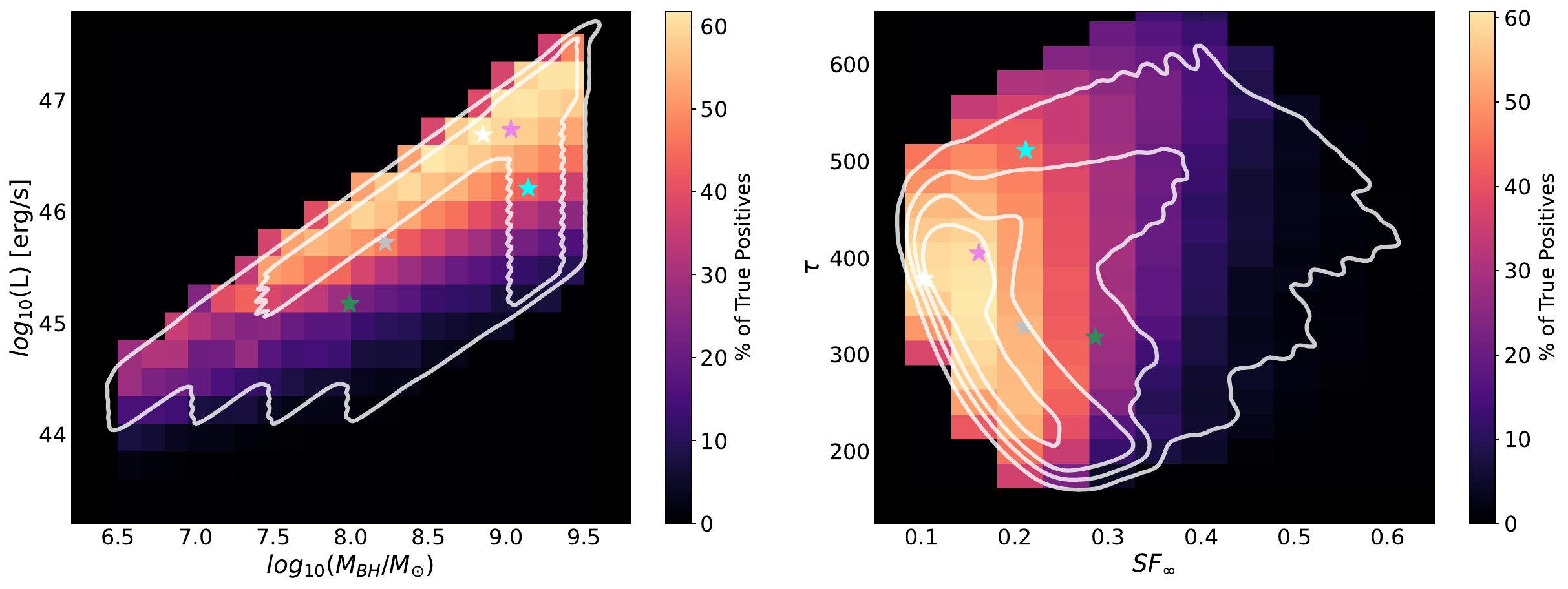}
\caption{2-D histograms of the true positive candidate demographics, contours (white) of the full binary population, and stars of five example binary candidates from Table \ref{tab:Candidates}. \textit{Left:} The luminosity and mass grid of true positives. We find that the most luminous and massive have the highest rates of true positive detection.  \textit{Right:} True positives in terms of DRW parameters, damping timescale $\tau$ and DRW amplitude \SFinf. Light curves made of small values of the DRW parameters report 60\% true positive detection rates. We see similar trends in Figure \ref{fig:FPDemographics} for the false positive population. }\label{fig:TPDemographics}
\end{figure*}

If we apply the same definition of binary detection to our true binary light curves, we find that only 38\% of all binary light curves are considered true positives. 
The low true-positive rate is a consequence of requiring \redchi $< 5.0$: about half of the binary light curves do not have best-fit binary models below this threshold, as shown in Figure \ref{fig:LCchi} and discussed in Section 3.
Figure \ref{fig:TPDemographics} depicts the demographics of the true positive population. The true positive detection rates peak at about 60\% for high-mass and high-luminosity quasars that have small DRW parameters. These trends mirror that of the false positive detections for this method. We also emphasize that a non-negligible number of binary light curves produce fit periods near the light curve baseline of five years, as seen in the right panel of Figure \ref{fig:NonbinDemo}. Most of these are binaries with ``weak" binary amplitudes ($A < 0.5$), with small \SFinf and over the full range of potential $\tau$ and binary periods.

\subsection{Recommendations for Rubin}\label{sec:RubinRecs}

We recommend exercising caution when using sinusoidal fits for binary SMBH detection. The very nature of pipeline-style sine fits and binary signals (in combination with the natural noise properties of the quasars we observe), will result in false positive detections. Even with analysis methods that model the quasar variability, this point holds \citep{Witt_2022}. If a sinusoidal fit is to be applied to quasar light curves in search of periodic signals, we can generally rely on the period and phase being accurate to within 10\%. This is especially true for binary signals that are strong, periods that are many times the damping timescale of the variability, and for luminous quasars that have smaller damping timescales (less massive) or structure functions (more massive). The more luminous a quasar is, the more likely it is to be a false positive detection due to its inherent variability properties. Up to 60\% of these quasars return false positive results across various black hole masses.

If the returned fit amplitude is larger than 0.4~mag, it's more likely to be a true binary than a false positive, as discussed in the previous subsection. There is large uncertainty associated with amplitude parameter recovery and the true underlying amplitude will usually be smaller than the best-fit value. Especially for weaker amplitude binary signals, the DRW amplitude strongly influences the fit and parameter recovery. 

We see some correlation with DRW-only fit periods and length of light curves as discussed in Section \ref{sec:Nonbins} and seen in the bimodal distribution of single quasar fit periods in Figure \ref{fig:NonbinDemo} (left panel, the white dashed line denotes our five-year baseline). If the best-fit period for a five-year light curve is near five or two and a half years, then it should be treated with caution and as a potential false positive that needs further analysis.

We can apply our recommendations to current binary SMBH candidates, with the caveat that they are candidates from different surveys, done using analysis methods that account for quasar variability. LSST is a higher-quality survey, so it would produce the lower limit of false positives. An example of the application of these recommendations is to attempt to apply them to long-reigning binary candidates like PG1302-102 \cite{Graham2015}. Assuming that when observed with Rubin/LSST, its calculated mass of log$_{10}$($M_{BH}/M_{\odot}$) = 8.3-9.4, its luminosity near its Eddington limit, and binary amplitude of 0.14 mag from its CRTS data would remain relatively unchanged, this candidate would be flagged as a likely false-positive candidate. Table \ref{tab:Candidates} contains four more examples of candidates in addition to PG1302-102 and our recommendation application, these candidates are plotted as stars on all panels on Figures \ref{fig:NonbinDemo} and \ref{fig:FPDemographics}.
It is notable that none of the best-fit CRTS candidates have amplitudes above our threshold $A>0.4$ for lower false-positive rates.
We note that \citet{Graham2015} used auto-correlation functions as well as sinusoid fits to identify the binary candidates shown in Figures \ref{fig:NonbinDemo} and \ref{fig:FPDemographics}; our results demonstrate that fitting only by periodic sinusoids will result very high false-positive rates for binary candidates with similar properties.

\begin{table*}[t]
    \centering    
    \begin{tabular}{|c|c|c|c|c|} \hline 
    \textbf{Candidate} (marker color) & Amplitude [mag]&  Period [days] &  $log_{10}$ (M/M$_{\odot}$)  & $log_{10}$ (L$_{bol}$/[ergs/s])\\ \hline 
     PG1302-102 (white) &  0.14 &  1, 884 $\pm$ 88 & 8.3-9.4 &  46.694   \\ 
     SDSS J143621+0727 (purple) &  0.08 &  1886 &  9.03 & 46.740  \\
     SNU J13120+0641 (cyan) &  0.09 &  1492 &  9.14 &  46.213  \\ 
     SDSS J080648+1840 (green) &  0.2&  892 &  7.99 & 45.173 \\ 
     SDSS J114749+1631 (gray) &  0.15 &  1449 &  8.22 &  45.726  \\ 
     \hline
    \end{tabular}
    \caption{Candidates from \cite{Graham2015} that approximately fall within our parameter space. The candidate amplitudes are estimated values from their respective light curves. The marker colors of the candidates are used in Figures \ref{fig:NonbinDemo}, \ref{fig:FPDemographics}, and \ref{fig:TPDemographics}. }
    \label{tab:Candidates}
\end{table*}

\subsection{Connections to Multi-Messenger Astrophysics}\label{sec:MMA}

An exciting prospect of electromagnetic detections of binary SMBHs are their connection to gravitational wave detection and multi-messenger astrophysics.  Our work shows that the binary phase and period are well-recovered in LSST DDF light curves.  Our work also shows a high false-positive rate for photometric detections of binary SMBH, thus they are not sufficient on their own. Rubin will be excellently poised for follow-up for binary candidates found via gravitational waves and in other wavelengths across the electromagnetic spectrum.

Our results are complementary to constraints on binary candidate populations. Casey-Clyde et al. 2023 (submitted) has shown that the most luminous candidates are also likely false positives. We also find that high luminosity quasars can mimic binary SMBHs when fitting with sine waves. 
This places more tension on the current lists of candidates, in which high luminosity sources are well-represented. Despite this tension, we should continue to observe these quasars as binary candidate population of N quasars goes as  $N \approx T_{obs}^{8/3}$.

We can further connect our recommended amplitude threshold of 0.4 mag to examples of physical binary parameters and to gravitational wave sensitivity with Pulsar Timing Arrays (PTAs). The change in apparent magnitude due to Doppler boost variability is directly tied to the line-of-sight velocity of the secondary black hole. By following the prescriptions laid out in \cite{Charisi_2022}, we can determine that a secondary black hole would need to be moving between 6.5\% and 10\% of the speed of light to produce that binary amplitude. These lower and upper bounds are determined by the redshift of the candidate. With a period detectable within a five-year LSST light curve (e.g., periods of 0.5 to 5.0~yr), these binary candidates could have a total masses between $10^{9.0}$ and $10^{10.6}$ $M_{\odot}$.
With a period, distance, and total mass, we can determine if these candidates would lie within current and future PTAs sensitivity ranges. Figure 6 of \cite{Charisi_2022} is an example of the overlap of EM and GW binary SMBH parameter space. 
Binaries with an orbital period between 180 and 1,825 days and masses of $10^{9.0}$ and $10^{10.6}$ $M_{\odot}$ would fall within potential PTA coverage. 
Keeping all other parameters equal, a larger change in apparent magnitude corresponds to a larger line-of-site velocity and a larger total binary mass, inferring higher sensitivity to the GW emission.

\section{Conclusion and Future Work}\label{sec:Conclusion}
 We have presented our work that included the generation and sine fitting of over 3.6 million Rubin/LSST quasar light curves, many of which contain binary SMBH signals. 
 We conservatively estimate that over 40\% of isolated, single quasars will result in a false positive detection of a binary SMBH system with a simple sinusoidal fit. The most luminous quasars in our study are the most likely candidates for a false positive detection with this method, with the false positive rate of these objects, specifically, being over 50\%. They are also, conversely, the binary types that produce the best recovery of our input parameters, as the structure function of the variability is low for these types of quasars. Less massive, less luminous quasars are more likely to dilute a binary signal because of their variability properties (namely large DRW amplitudes). 
 
 From our work, we provide the following recommendations for Rubin/LSST binary SMBH searches and identification:

 \begin{itemize}
     \item Periods and phases of  binary signal can reliably be recovered within 10\% of the input binary signal value via sinusoidal curve fits roughly 70\% of the time, separately, or 62\%, together.
     \item Binary amplitude is the least-reliable parameter to recover, but best-fit amplitudes larger than 0.4 magnitudes can be used to boost confidence in a binary candidate.
     \item Best-fit periods that are near the length or half-length of their respective light curves should be treated with caution.
     \item Sources that are massive and luminous are more likely to be a false positive than a real binary if the DRW is not accounted for.
 \end{itemize}

\subsection{Future Work}

The purpose of this paper was to explore a the detectability of binary SMBHs for a representative quasar population in Rubin/LSST DDF observations.
In the future, the light curve generation could be adapted to use
parameter distributions
that are more directly tied to observed quasar population and/or limits on binary SMBHs inferred from current gravitational wave searches.

Beyond light curve generation, there are more analysis methods we plan to apply to the light curves, like Lomb-Scargle periodograms and Auto-Correlation Functions (\citealp{Lomb}; \citealp{Scargle}; \citealp{Lombscargle}; \citealp{EdelsonKrolik1988}; \citealp{McQuillan2013ACFs}) and machine learning algorithms. We hope to revisit this analysis with the development and inclusion of computationally inexpensive DRW fits, in addition to the sine fits. This 
could result in more effective triage of false positives. There is more room to explore how these results differ with different cadences (like main LSST survey WFD cadence versus the DDFs) and longer baseline light curves (the full 10-year LSST baseline and beyond). 

We also plan to apply these analysis pipelines to existing large-scale surveys, like the Catalina Real-time Transient Survey, as done in \cite{Graham2015} and \cite{Witt_2022}.
Our methods can also be applied to make predictions for
other upcoming time-domain surveys, like the High Latitude Time Domain Survey of the Nancy Grace Roman Space Telescope (Roman) which is expected to launch in 2027 \citep[at the time of writing this paper][]{spergel2013ROMAN, spergel2015ROMAN}. Roman may also be able to detect massive binary systems through its observation of $10^4$ - $10^6$ quasars \citep{haiman2023massiveROMAN, shen2023Roman}.

\acknowledgments
JRT, MCD, KEG, and AH acknowledge support from NSF grant CAREER-1945546. MCD acknowledges support from a NSF Graduate Research Fellowship. LB acknowledges support from Cottrell Scholar Award \#27553 from the Research Corporation for Science Advancement and from NSF award AST-1909933. WNB acknowledges financial support from NSF grant AST-2106990. JACC is
supported by the National Science Foundation’s NANOGrav
Physics Frontier Center, Award Number 2020265 and by the National Science Foundation under Grants
NSF PHY-2020265, and AST-2106552. MC and JCR acknowledge support from NSF award AST-2007993. 
The authors wish to thank London E Willson and Chiara MF Mingarelli for their helpful discussion and insight. MCD wishes to especially thank Dani Lipman and Hugh Sharp for their support.

\vspace{5mm}
\software{ astroML \citep{astroML}, SciPy \citep{2020SciPy-NMeth}, NumPy \citep{NumPy}, LMFits \citep{lmfits}}

\bibliography{binary}{}
\bibliographystyle{aasjournal}

\end{document}